\documentclass[preprint,a4paper,3p,10pt,times,twocolumn,dvipdfm]{elsarticle}
\usepackage{graphics}
\usepackage{graphicx}
\usepackage{epsfig}
\usepackage{amssymb}
\usepackage{amsthm}
\usepackage{natbib}
\usepackage{latexsym}
\usepackage{mathrsfs}
\usepackage{amsmath}
\usepackage{amscd}
\usepackage{color}
\usepackage{verbatim}

\journal{Phys. Lett. A}

\begin{document}

\begin{frontmatter}

\title{Conservative Generalized Bifurcation Diagrams}

\author[udesc]{Cesar Manchein}
\ead{cmanchein@gmail.com}
\author[ufpr]{Marcus W. Beims}
\ead{mbeims@fisica.ufpr.br}

\address[udesc]{Departamento de F\'\i sica, Universidade do 
Estado de Santa Catarina, 89219-710 Joinville, Brazil} 
\address[ufpr]{Departamento de F\'\i sica, Universidade Federal 
do Paran\'a, 81531-980 Curitiba, Brazil}

%

\begin{abstract}

Bifurcation cascades in conservative systems are shown to exhibit a 
{\it generalized} diagram, which contains all relevant informations regarding 
the location of periodic orbits (resonances), their width (island size), 
irrational tori and the infinite higher-order resonances, showing the 
intricate way they are born. Contraction rates for islands sizes, along 
period-doubling bifurcations, are estimated to be $\alpha_I\sim 3.9$. 
Results are demonstrated for the standard map and for the continuous 
H\'enon-Heiles potential. The methods used here are very suitable to 
find periodic orbits in conservative systems, and to characterize the 
regular, mixed or chaotic dynamics as the nonlinear parameter is varied.
\end{abstract}

\end{frontmatter}

\section{Introduction}
\label{Introduction}

Almost all physical systems in nature are so complex that long
time predictions are nigh impossible. This is a characteristic 
of nonintegrable chaotic systems whose effects are visible in 
celestial mechanics, plasma physics, general relativity, quantum 
physics, communications problems, hard beats, social and stock 
market behaviors, weather forecast, among others. Thus, the 
precise description of the dynamics in nonintegrable systems is 
essential for the understanding of nature. One intrinsic and
fundamental phenomenum of such nonintegrable systems, 
is that realistic stable orbits may vanish (or be born)
when the nonlinear parameter varies. This can lead to a
complicated behavior with a cascade of new orbits, which is 
satisfactory described by a bifurcation diagram \cite{lichtenberg92}. 
One example is the period doubling bifurcation (PDB) cascade, well 
understood in one dimensional dissipative discrete systems
containing one parameter, where the intervals in the parameter, 
between successive PDBs, tend to a geometric progression with an 
universal ratio of $1/\delta = 1/4.66$, with $\delta$ being the 
Feigenbaum constant. For two dimensional dissipative systems, 
bifurcation cascades manifest themselves inside periodic stable
structures in the two parameter space, which appear to be generic 
for a large class of systems. 
Shrimp-like structures are one example, and appear in the 
dissipatives H\'enon \cite{jasonPRL93,grebogi93} and ratchet 
\cite{alan11-1} discrete systems, in continuous models 
\cite{bonattoR07}, among many others. Such structures allow 
an analysis of geometric approximation ratios, and provide 
a very clear understanding of the dissipative dynamics. 
In {contrast to dissipative systems, for} conservative 
nonintegrable systems the description of bifurcations cascades 
is much more complicated. 
It is known \cite{lichtenberg92,reichl04}, that by magnifying stable 
points in the phase space, a mixture of surrounding stable and 
unstable fixed points is found. This repeats itself for every 
stable fixed point,  as explained by the Poincar\'e-Birkhoff 
theorem \cite{ott-book}.  { How higher order stable orbits bifurcate, 
the islands around them vary and are interconnected, and irrational 
tori behave as the nonlinear parameter of the system changes,} is an 
interesting problem that still deserves 
to be deeply investigated. For $2$ degrees of freedom area-preserving 
discrete systems, the intervals in the parameter between successive 
PDBs, tend to a geometric progression with a ratio of 
$1/\delta_H \sim 1/8.72$ \cite{lichtenberg92}. 
Results have also been extended to higher-dimensional systems (please 
see \cite{helleman81,helleman87,mao88,kim94} for more details).

This work uses convergence properties of the Finite Time Lyapunov 
Exponent (FTLE) to explore the dynamics of conservative systems
in a mixed plot: initial condition (IC) {\it versus} the nonlinear 
parameter. The location of stable orbital points is easily found 
in such plot, and conservative bifurcation diagrams recognized 
to have a generalized form, containing  infinite sub-diagrams
with all rational/irrational tori from the periodic orbits (POs), 
independent of the period. This is a nice complete description, 
and extension, from an early work \cite{feigen81}, done for another 
dynamical system. Results are remarkable and show the very 
complex, self-similar and generic bifurcation structure in conservative 
systems {with only one parameter.  They also} suggest 
that contraction rates  for the islands sizes, along PDBs, approach the 
constant 
$\alpha_I\sim 3.9$. A detailed numerical analysis is performed for 
the standard map, including the use of the first recurrence times 
instead the FTLEs, to demonstrate that results are independent of 
the method. A generalization is shown for the continuous 
H\'enon-Heiles potential.

\section{Discrete Model: The Standard Map}
\label{model}

{To start it is appropriate to present results using a well
 known general model with wide applications, the Chirikov 
standard mapping, which is given by \cite{chirikov79}: 
\begin{eqnarray}
  \left\{
\begin{array}{ll}
p_{n+1} = p_n + (K/2\pi)\sin(2\pi x_n)\quad \mbox{mod}~1, \cr
x_{n+1} = x_n + p_{n+1}\quad  \mbox{mod}~1.\cr
\end{array}
\right.
\label{map}
\end{eqnarray}
$K$ is the nonlinearity parameter and $x_n,p_n$ are
respectively position and momentum at discrete 
times $n=1,2,\ldots,N$. It is known that { 
period-$1$ (shortly written per-$1$)} fixed 
points are $p_1=1/2 m$ ($m$ integer) and $x_1=0,\pm 1/2$. 
The point $x_1=0$ is always unstable while $x_1=\pm 1/2$ 
becomes unstable for $K>4$. There exist also per-$1$ 
fixed points related to accelerator modes 
\cite{chirikov79} whose stability condition is 
$|2\pm K\cos{x_{1l}} |<2$, with $K\sin{x_{1l}}=2\pi l$ and $l$ 
integer. For higher periods there are primary families of 
periodic points (which exist in the limit $K\to 0$) and 
bifurcation families which are born only for larger values of $K$
(See \cite{lichtenberg92} for more details). 

\section{Method: Finite Time Lyapunov Exponents}
\label{method}

The key idea for the success of our proposal is the observation 
that the numerical  convergence of the FTLEs is {\it distinct} 
for different ICs (same nonlinear parameter $K$), even between 
regular trajectories. To make this clear, consider, for example, 
{ that the IC is exactly on the stable (for $K\le 4$) fixed
point $x_0=1/2$. It can be calculated analytically that the 
corresponding Lyapunov exponent (LE), after one iteration, is 
exactly zero. But now consider that we start with an IC close
to this fixed point, say $x_0^{\prime}=1/2+\Delta x_0$, which can be 
an irrational regular torus close to $x_0$. Assuming that 
$(\Delta x_0)^2\approx 0.0$, the standard map can be linearized 
around  $x_0^{\prime}$  and the FTLE determined analytically,
after $n$ iterations, from the eigenvalues of the composed Jacobian 
$\{(1,-K),(1,1-K)\}^n$. This FTLE is plotted in Fig.~\ref{LExX0}(a) 
as a black continuous line, and it converges exactly to zero only 
when $n\to\infty$. This means that ICs from stable tori around the 
fixed point, take a longer time to converge 
exactly to zero than the fixed point itself.} This behavior can 
also be observed numerically by { determining the FTLEs 
using Benettin's algorithm \cite{benettin80,wolf85}, which includes 
the Gram-Schmidt re-orthonormalization procedure.} We use the six 
exemplary orbits shown in the inset 
of Fig.~\ref{LExX0}(a): the stable fixed point, demarked with a cross, 
and the five irrational tori around the fixed point. All  
trajectories are regular and have LEs exactly equal zero for
infinite times. However, when calculating the FTLEs 
for the distinct ICs, we observe that ICs closer to the fixed point 
converge faster to zero. In Fig.~\ref{LExX0}(a), the decay curves for 
the FTLEs {\it versus} times are plotted 
(starting from below) for the six ICs shown in 
the inset (starting from the fixed point). As ICs 
are taken more away from the periodic point, FTLEs converge slower 
to zero. The magnitude of the FTLEs between different 
irrational tori are really small and not significant for any 
purpose. However, for the IC exactly on the PO, the FTLE converges 
faster to zero than other ICs around it. { Essential to mention 
is that this is not a numerical convergence artefact due to the 
numerical method, but an analytical property, as shown 
above (continuous line in  Fig.~\ref{LExX0}(a)).}
\begin{figure}[htb]
  \centering
  \includegraphics*[width=0.55\linewidth]{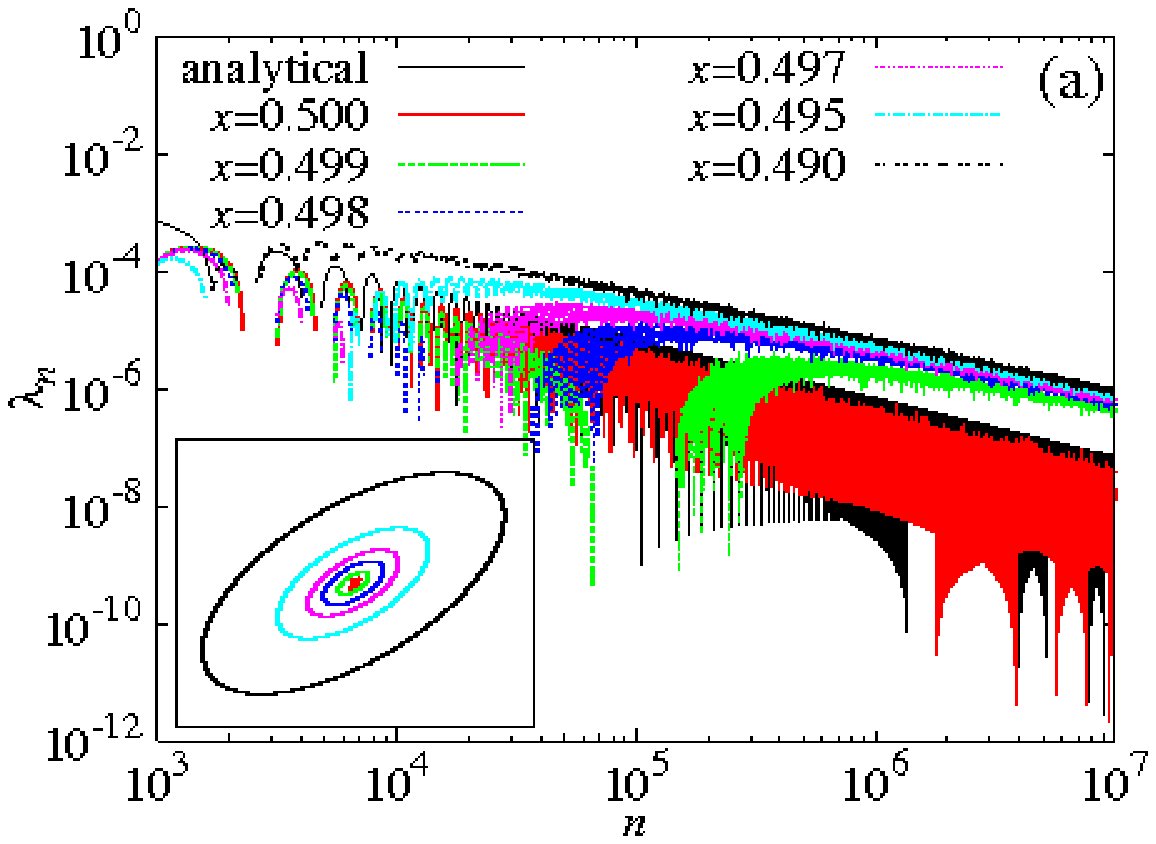}
  \includegraphics*[width=0.4\linewidth]{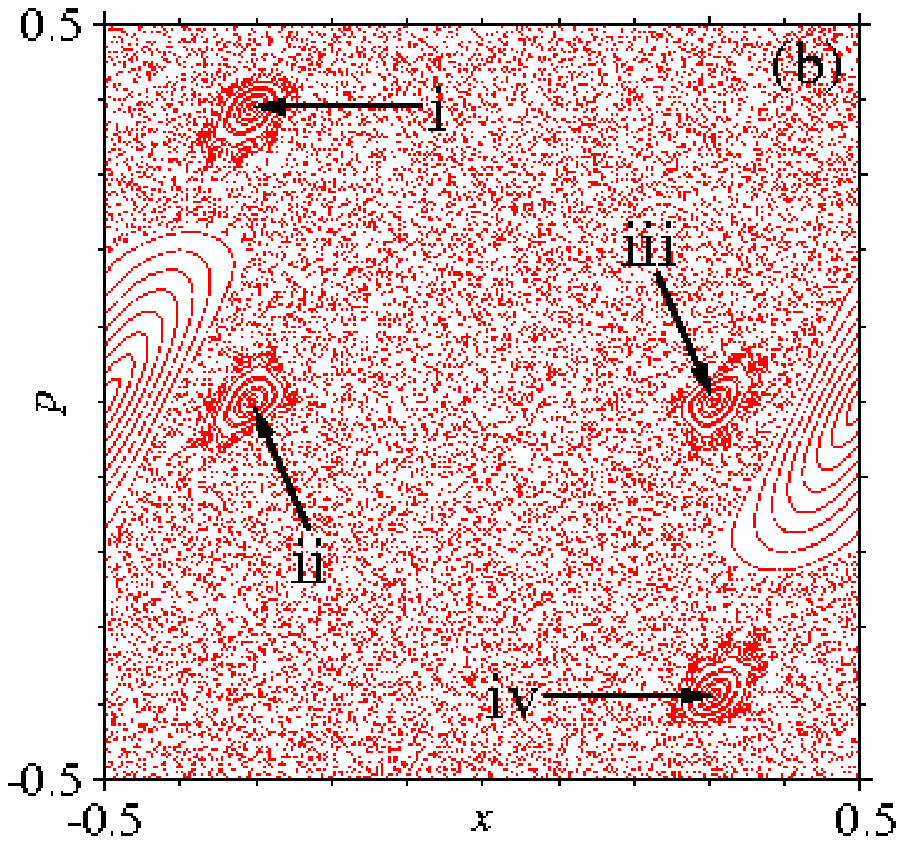}
 \caption{(Color online) (a) Log-log plot of the FTLEs (starting 
from below) as a function of the iteration time for the six 
trajectories shown in the inset (starting from the fixed point), 
and (b) phase space for $K=2.6$.}
  \label{LExX0}
\end{figure}
Thus, even though such small FTLEs are insignificant to distinguish 
between the irrational tori, they allow us to recognize 
where POs are {\it located} in phase space, and to understand the 
very complex and self-similar behaviors, which occur close to the 
POs as the nonlinear parameter $K$ changes. For larger times, 
FTLEs in Fig.~\ref{LExX0}(a) continue to decrease linearly to zero, 
until the machine precision is reached, and when the actual method 
cannot be used anymore. { Even though the method was 
explained using the standard map, it is equally applied to other 
conservative dynamical systems.}

\section{The Generalized Diagrams}
 
{Using the properties explained in Section \ref{method}}, 
a very clarifying plot can be 
constructed, which allows us to recognize the bifurcation diagram
in conservative systems in a simple way. Figure \ref{XxK} shows 
the FTLE (see colors bar) in the mixed space $x_0\times K$ with $p_0=0.0$.
Black lines are related to those ICs for which the FTLEs converge 
faster to zero and POs exist (this was checked explicitly for many 
black lines). Dark to light yellow points, around the main black lines, 
are related to irrational tori and also define the size of the 
corresponding island. These are regular trajectories for which the
associated FTLEs are still not that close to zero, as those FTLEs 
from ICs which start exactly on the black lines (POs). For $n\to\infty$ 
these points also become black. Light  yellow, blue to red points, are 
ICs related to chaotic trajectories. 
\begin{figure}[htb]
  \centering
  \includegraphics*[width=0.9\linewidth]{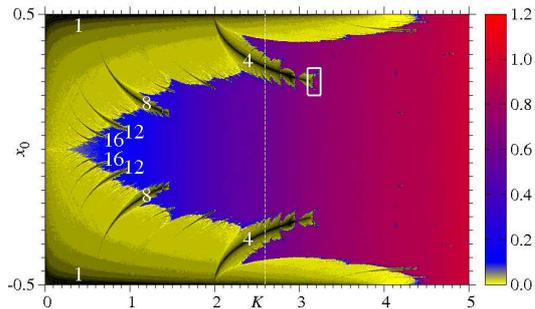}
 \caption{(Color online) FTLEs in the mixed space 
$x_0\times K$ for the standard map with $p_0=0.0$, a grid of
$10^3\times 10^3$ points and $10^4$ iterations.}
  \label{XxK}
\end{figure}
Clearly we observe that the fixed point $x_0=x_1=0.0$ is unstable 
(positive FTLEs along the line  $x_0=0.0$) for any $K$ value. 
In the chaotic region, FTLEs 
increase monotonically with $K$ and the dynamics is not much 
exciting. However, in the regular to chaotic transition
region, trajectories combine themselves in a very complicated way,
bifurcations occur and the dynamics is very rich as $K$ changes. 
For clarification, Fig.~\ref{LExX0}(b) shows the corresponding phase 
space dynamics for $K=2.6$ [see white dashed line in 
Fig.~\ref{XxK}]. One per-$1$ (at $x=\pm 0.5, p_0=0.0$) and one 
per-$4$ (see numbers i, ii, iii, iv showing the orbital points) 
POs are shown. The per-$1$ orbital points correspond to 
the black points in Fig.~\ref{XxK} located at $x_0=\pm 0.5$ and 
$K=2.6$. This is the primary resonance. The two orbital points 
(ii,iii), from the per-$4$ PO, belong to the black curves in 
Fig.~\ref{XxK}. Since we always 
started along the initial line $p_0=0.0$, Fig.~\ref{XxK} is able 
to detect only the two orbital points (ii,iii) along this line. The
whole dark to yellow region around $x_0\pm 0.5$ (for $K\le 4$)
in  Fig.~\ref{XxK}, gives the island size and is related to all
irrational tori around this PO.

ICs which leave to stable POs (black lines), are characterized by 
the ``Christmas tree'' structure around it. For example, for 
$K=2.0$ in Fig.~\ref{XxK}, two per-$4$
trees are born, one at $x_0=-0.5$, and the other one at $x_0=0.5$.
All trees have a main ``stem'' in the center, which is black 
and gives the location of the stable PO when $K$ varies.
Since both mentioned trees are born from $x_0=\pm 0.5$, they 
are secondary resonances with per-$4$. {A sequence of identical 
trees (secondary resonances) with smaller and smaller sizes, and 
increasing periods $8,12,16,\ldots$ [see numbers on the sequence
of trees Fig.~\ref{XxK}], are observed by decreasing $K$. The 
stem of these trees, which are all born from the PO $x=\pm 0.5$, 
are related to the rational tori. The width of the tree is 
the width of the island along the line $p_0=0.0$. In addition, 
these trees approach to each other as they get closer to the 
unstable fixed point at $x_0=0.0$ and $K\to 0.0$. A large 
sequence of approximating trees is also observed (not shown 
here) close to $K\sim 4.0$ (where the per-$1$ PO becomes 
unstable) and $x_0\to \pm 0.5$. 
\begin{figure}[htb]
  \centering
 \includegraphics*[width=0.48\linewidth]{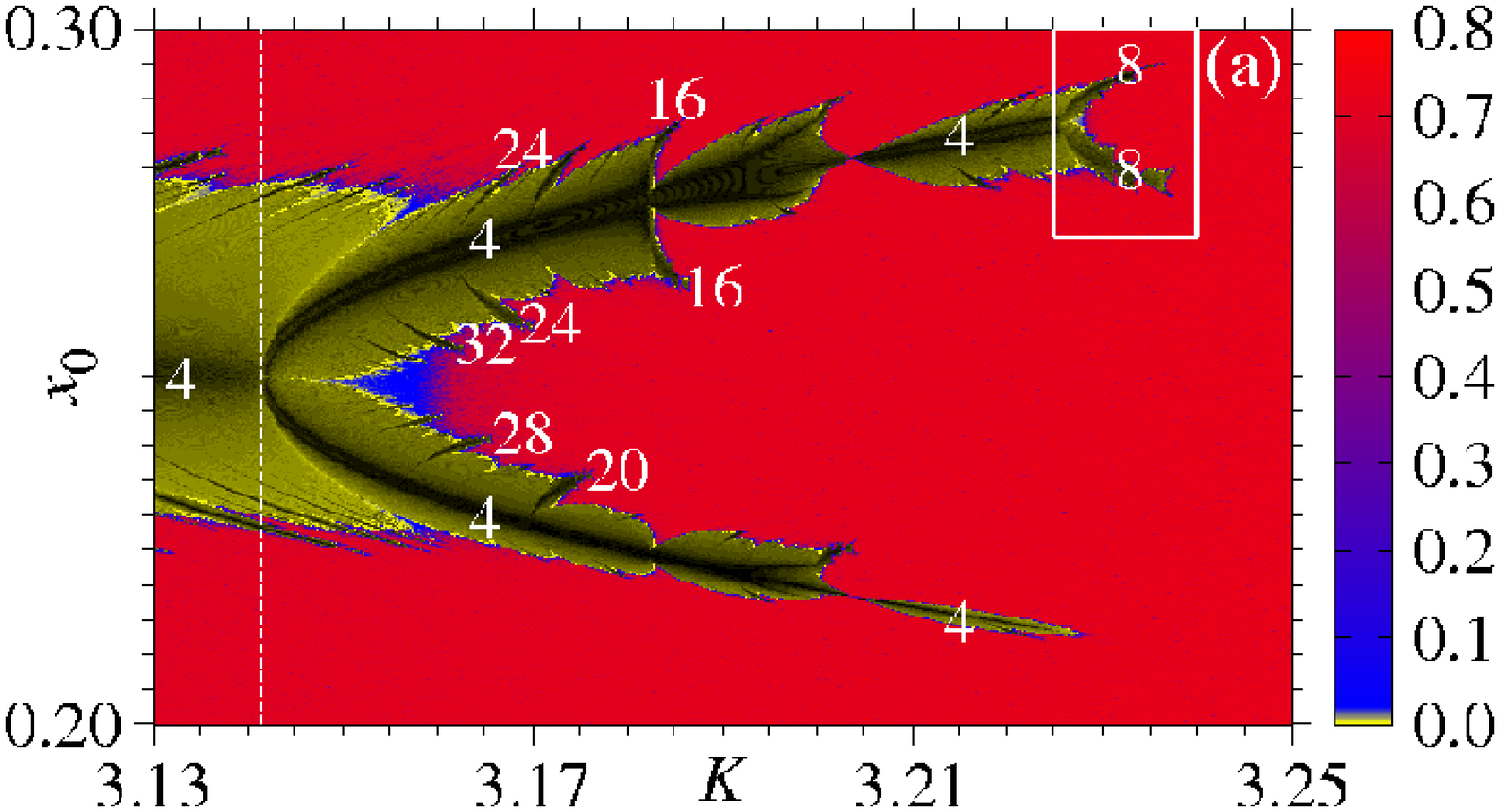}
 \includegraphics*[width=0.48\linewidth]{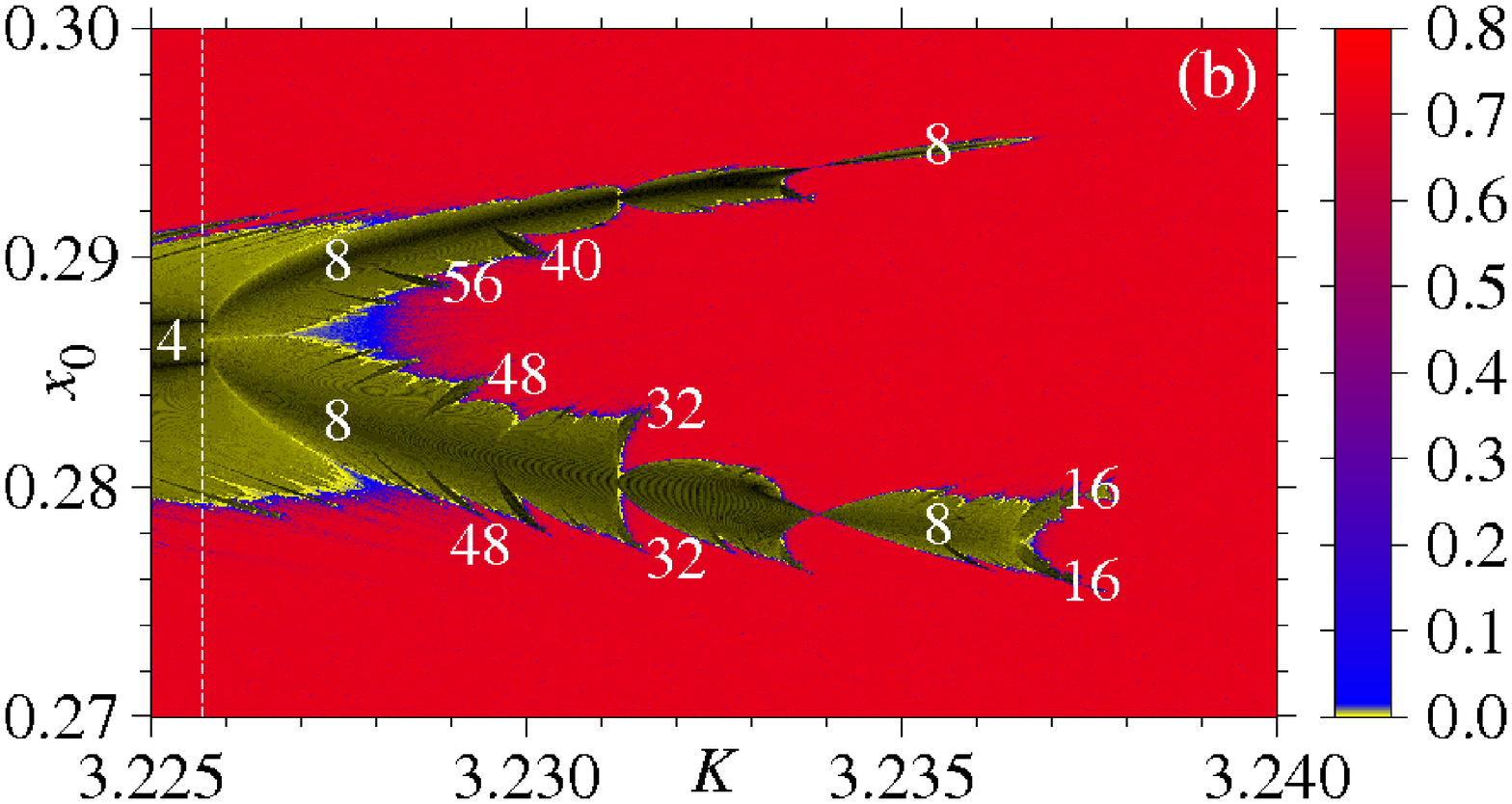}
  \caption{Magnifications of (a) the per-$4$ fork from 
the box from  Fig.~\ref{XxK} and (b) per-$8$ fork from box 
from (a). Numbers indicate the periods of the POs.}
  \label{zoomP4}
\end{figure}
The main stems in the above trees exist until they become 
unstable at the very end of the tree (on the right), where a 
pitchfork bifurcation occurs. This is better visualized in 
the magnification shown in Fig.~\ref{zoomP4}(a) which,
roughly speaking, summarizes results contained in Figs.7 and 8 from 
\cite{feigen81}, obtained for another dynamical system. However,  
Fig.~\ref{zoomP4}(a) contains many essential additional informations 
about the rich underline dynamics, as described below. 
The main stem (stable PO) becomes unstable close to 
$K\sim 3.145$ and $x_0\sim0.25$, as can be recognized by the yellow 
points (see dashed line). Simultaneously around this point, two black 
lines (stable POs) are born with the same per-$4$, 
characterizing the pitchfork bifurcation. In fact, a ``fork''-like 
structure (diagram) is born at the bifurcation point and contains 
two new main trees. This fork is the {\it generalized diagram}, which 
lives in the {\it entire} portion of the mixed plot where the 
{\it transition} from regular to chaotic motion occurs. This was 
checked for many other magnifications, even for higher order periods. 
Since the interesting dynamics and all bifurcation processes occur 
through the fork and it trees, it is enough to discuss just one fork 
to understand the very rich dynamics from the whole Fig.~\ref{XxK}.

For this we start discussing the magnification of the per-$4$ 
fork shown in Fig.~\ref{zoomP4}(a). Inside one fork, two main trees
always exist: one called the Line Tree (LT), where PDBs 
occur for ICs {\it on} the line $p_0=0.0$; and the other 
is the Outline Tree (OT), where PDBs occur for ICs {\it outside} 
$p_0=0.0$. In the example from  Fig.~\ref{zoomP4}(a), the 
OT is the 
lower one and the LT is the upper one, which is identical to
all secondary trees shown in  Fig.~\ref{XxK}. LT and OT
have a main stem related to {\it distinct} POs, but with the 
{\it same} period [see the numbers in Fig.~\ref{zoomP4}(a)]. 
Emerging almost perpendicularly from the main stems, 
infinite smaller sub-trees are born below and above. Such new 
sub-trees have also a main stem related to another PO with higher 
period. In the ramifications shown in Fig.~\ref{zoomP4}(a), the new 
POs emerging from the LT, as $K$ decreases, are: $8$(new fork, see 
box)$,16,24,32,\ldots$  while for the OT are: $20,28,36,\ldots$. 
These new ramifications (sub-trees) are the rational tori 
(higher-order resonances) which live around the per-$4$ point. 
At the end of the fork (on the right), the main stems from 
the LT and OT suffer a PDB phenomenum. This is shown in details
in Fig.~\ref{zoomP4}(b), which is a magnification of the 
white box per-$8$ fork from Fig.~\ref{zoomP4}(a). 
Besides the new trees emerging from the main stem, a continuous 
transition to lighter colors occurs when we go away from the 
main stem, in the direction of the border of the tree. All
these points are related to irrational tori around the main
PO. At the border of all trees, the breakdown of the last irrational 
torus, of the corresponding PO, can be observed by the 
{transition from yellow} to red points, where positive 
FTLEs appear and the dynamics becomes chaotic. 
Remarkable is that the dynamics around this new unstable point, 
as $K$ increases, is similar to the dynamics around the 
unstable per-$1$ ($x_0=0.0$) orbit, observed in 
Fig.~\ref{XxK} for $K$ close to zero. 
Collecting the main properties of 
the generalized fork, suppose it has a per-$q$ for the main 
stems of the LT and OT. The ramifications have distinct periods, 
namely for the LT we have, starting from the right border of the fork: 
$2q$ (new fork), and the ramifications related to the rational tori 
have periods $4q,6q,8q,\ldots$. For the OT we have $2q$ (not seen 
along the line $p_0=0.0$) with the ramifications $5q,7q,9q,\ldots$.
From successive magnifications (not shown) of the generalized 
fork, we found that it contracts with a ratio $\delta_F\sim 8.7$,
and that the islands sizes, at the PDBs, approach the estimated 
contraction rate $\alpha_I=d_{q}/d_{2q}\sim 3.90\pm0.04$, where $d_q$ is 
the island size for $q=128$. {
In order to estimate $d_q$, we magnified the forks with main 
periods $q=128$ and $256$,  using times $n=10^?$.
For such times, the border line between the last irrational tori 
around the POs and the chaotic motion becomes evident, and $d_q$ 
can be determined from the numerical data of the FTLEs.}

\section{A continuous model}
\label{henon}

The general statement of our results can be confirmed showing 
the existence of trees in the continuous H\'enon-Heiles (HH) potential 
$V(x,y)=(x^2+y^2)/2 +x^2y - y^3/3$. It is a truncation 
of the Toda potential \cite{lichtenberg92} used in celestial 
mechanics and extensively {mentioned as a benchmark} 
to nonlinear conservative dynamics.
\begin{figure}[htb]
  \centering
 \includegraphics*[width=0.9\linewidth]{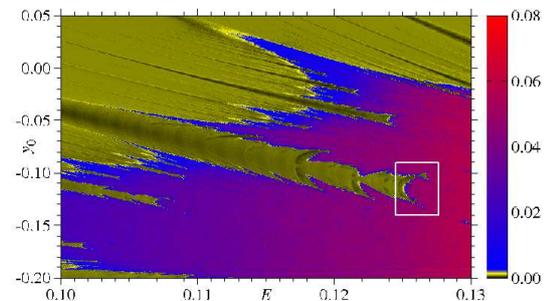}
 \caption{FTLE in the space $E\times y_0$ for the HH potential
with $x_0=p_{y_0}=0.0$, while $p_{x_0}$ is obtained from energy 
conservation.}
  \label{HH}
\end{figure}
In this case the nonlinear parameter is the total energy
$E$. For $E=1/24$ the dynamics is regular, and when $E\to 1/6$
it becomes mixed and finally chaotic. Figure \ref{HH} shows 
the FTLEs in a portion of the the mixed plot $E\times y_0$ for 
the HH problem. As in Fig.~\ref{XxK}, black lines are related 
to those ICs which leave to close to zero FTLEs, where POs are 
present. Dark to yellow  colors are related to the rational/irrational 
tori. Clearly we observe the existence of the stable trees, the forks 
(see white box for one example) and the infinite sequence of
trees ramifications (higher order resonances), always appearing in 
form of the tree described in details for the standard 
map.

\section{Changing the method: First recurrence times}

The purpose of this section is to show that the generalized diagrams 
can be obtained using another method. Despite results using the 
FTLEs are totally trustworthy, it is proper to use a distinct method. 
For this we reproduce Fig.~\ref{XxK} using the first recurrence 
times (FRTs) instead the FTLEs. We start the simulation with ICs 
inside the interval $-0.5<x_0<0.5,\, p_0=0.0$, iterate the standard 
map (\ref{map}) and reckon $\Delta x_n = x_0-x_{n}, \Delta p_n = p_0-p_{n}$. 
When $|\Delta x_n|<10^{-4}$ and $|\Delta p_n|<10^{-4}$, we record $T_1=n$.
For $n> T_{max}=10^4$ we stop the simulation. In this way we compute 
the number of iterations $T_1$, which a given trajectory needs to 
come back  to the initial point (inside a circle of radius $10^{-4}$)
 for the first time.  It is expected that for POs the
time $T_1$ is smaller (equal to the period of the orbit), and that
for chaotic trajectories the limit $n= T_{max}$ is reached. In between,
all values of $T_1$ are possible, and are usually related to irrational 
tori or POs with very large periods.
\begin{figure}[!ht]
    \centering
  \includegraphics[width=0.9\columnwidth]{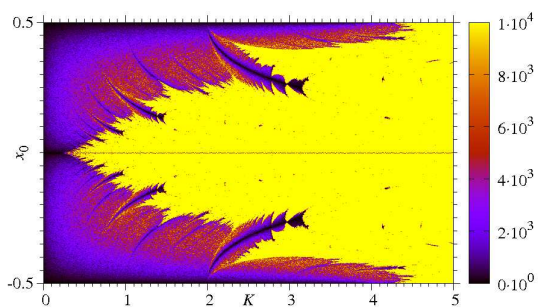} 
   \caption{FRT $T_1$ (see color bar) for the 
  standard map (\ref{map}) in the plot ICs {\it versus} K.}
  \label{time}
\end{figure}
Results are shown in Fig.~\ref{time}. Plotted is $T_1$ (see color bar)
as a function of ICs $x_0$ and K. Black points are related to smaller 
times and are identified as the POs. Purple to red points 
are the irrational tori with intermediate values of $T_1$. Yellow points 
are related to chaotic trajectories, which never returned (for 
$n\le 10^4$) to the IC. Even though Fig.~\ref{time} is plotted with 
different colors, all trees, forks structures, widths and scalings, are 
identical to Fig.~\ref{XxK}. In addition, the FRTs
are able to find the location of {\it unstable} POs, as for example, the 
straight horizontal line at $x=0.0$, which is the unstable fixed point 
for any $K$.

The reason why the FRTs work so well to reproduce the results from the 
FTLEs is a discussion which we cannot explore in a satisfactory way in 
this work. It has also other important properties.
The main point here is to demonstrate that our results 
are real physical effects, relevant and independent on the method: 
FTLEs, FRTs, or any other quantity whose time 
convergence is affected by the ICs. 

\section{Conclusions}
\label{conclusions}

To conclude, period doubling bifurcations in conservative systems are 
shown to exhibit a generalized stable diagram in the plot: initial 
conditions {\it versus} the nonlinear parameter. This diagram is 
composed of two main trees whose main stems are the periodic orbits 
(resonances) and infinite ramifications, which are all rational tori 
(higher-order resonances) surrounding the periodic orbits. Results are 
discussed in details for the standard map and extended to the continuous 
H\'enon-Heiles potential. The generalized trees were also confirmed to 
exist (not shown), for the nontwist standard map \cite{szezech09} and the 
hydrogen atom 
subjected to an external magnetic field \cite{beims-gallas00}, and can 
be recognized in \cite{feigen81}, giving additional evidences about 
their wide applicability in conservative systems. First recurrence 
times and finite time Lyapunov exponents were used for the standard map,
showing that results are independent of the method. Other methods could be 
used to find the trees as, for example, the finite time rotation numbers.
This is what Fig.~5, from the recent work about the nontwist standard map 
\cite{Szezech12}, suggests, providing one more example of a distinct 
system, with another method, where the generalized diagram should appear. 
Our results are also independent of the periods of the periodic orbits,
and provide general clues for conservative systems concerning: the 
regular/mixed/chaotic dynamics, the location of stable orbital points, 
the generic mechanism of bifurcation phenomenum, and that contraction 
rates of islands sizes, along period doubling bifurcations, follow  
$\alpha_I\sim 3.9$.
To finish, from the present analysis it is possible to conjecture that, 
for higher dimensional systems, the dynamics outside the generalized 
bifurcation diagram, and for small nonlinearities, should produce Arnold 
stripes \cite{cesar-marcelo12}, which display the ICs from the stochastic 
Arnold web as a function of the nolinearity parameter. 

\section{Acknowledgments}
The authors thank FINEP (under project CTINFRA-1) and MWB thanks CNPq 
for financial support. 

%

\end{document}